\documentclass[twocolumn,showpacs,preprintnumbers,amsmath,amssymb,A4paper,prd]{revtex4}

\usepackage{graphicx}
\usepackage{dcolumn}
\usepackage{bm}

\begin{document}

\preprint{APS/123-QED}

\title{On $\bm{v^2/c^2}$ expansion of the Dirac equation with external potentials}

\author{Wlodek Zawadzki}
 \affiliation{Institute of Physics, Polish Academy of Sciences\\
       Al.Lotnikow 32/46, 02--668 Warsaw, Poland\\
 }

\date{\today}

\begin{abstract}

The $v^2/c^2$ expansion of the Dirac equation with external
potentials is reexamined. A complete, gauge invariant form of the
expansion to order $(1/c)^2$ is established which contains two
additional terms, as compared to various versions existing in the
literature. It is shown that the additional terms describe
relativistic decrease of the electron spin magnetic moment with
increasing electron energy.

\end{abstract}

\maketitle

\section{\label{sec:level1}Introduction\protect\\ \lowercase{}}

A semirelativistic expansion of the Dirac equation ( up to
$v^2/c^2$ terms ) is treated in almost all sources on quantum
mechanics which include elements of the relativistic quantum
theory. This expansion is of importance for problems involving
electric and magnetic potentials for which exact solutions of the
Dirac equation do not exist. The important examples include
electron behavior in atoms, molecules and solids in the presence
of a magnetic field. Surprisingly, various versions of the
$v^2/c^2$ expansion quoted in the literature vary strongly from
one source to the other \cite{pp1}-\cite{pp20}. Almost all final
forms are incomplete, quite a few are not gauge invariant.

The purpose of this contribution is to establish the complete and
gauge invariant form of the $v^2/c^2$ expansion of the Dirac
equation with electric and magnetic potentials and to interpret
its physical content.

There exist two ways to carry the $v^2/c^2$ expansion. One is to
consider large and small components of the wave function for
electrons having positive ( or negative ) energies and to find the
large component to the desired order of $1/c$ by iteration. In
this method one has to take into account a changed normalization
condition for the large components ( to the same order ). The
other way is to use the Foldy-Wouthuysen unitary transformation in
order to eliminate odd operators in the Dirac equation to the
desired order. In this method the normalization condition is taken
into account automatically since a unitary transformation does not
affect the normalization. Both methods, if carried out
consistently, lead to the same result up to the order $(1/c)^2$ (
see de Vries \cite{pp12} ). In our procedure we use the unitary
transformation as it is more systematic.

 \section{\label{sec:level1}RESULTS AND DISCUSSION\protect\\ \lowercase{}}

The Dirac equation for an electron reads
\begin{equation}
H=\beta mc^2 + U + O\;\;,
\end{equation}
 where $m$ is the rest
electron mass, $U=-eV$ is the potential energy and $O=c\;{\bm
{\alpha}} \cdot \bf{\Pi}$ is the kinetic energy. In order to avoid
misunderstandings concerning the order of $1/c$, we put explicitly
$c$ in Eq. (1) and write the canonical momentum ${\bf \Pi} = {\bf
p}+e {\bf A}$, i.e. the vector potential appears without $c$ ( SI
units). Symbol $e$ signifies absolute value of the electron
charge. Matrices $\beta$ and $\bm{\alpha}$ are taken in the well
known Dirac form. We assume that the potentials $V$ and
$\textbf{A}$ do not depend on time.

The kinetic term $O$ represents an odd part of the Hamiltonian
mixing upper and lower components of the wave function. In order
to eliminate the odd part the following Foldy-Wouthuysen
transformation is introduced
\begin{equation}
H'=exp(iS)Hexp(-iS)\;\;,
\end{equation}
 where  $S=(-i/2mc^2)\beta O$ .The
exponentials in Eq. (2) are expanded into power series to the
desired order of $1/c$, taking into account that $S\sim 1/c$. The
choice of $S$ eliminates the odd term $O$ in Eq. (1), but
introduces other odd terms of higher order in $1/c$. These are
eliminated by two consecutive transformations $S'$ and $S''$
leading to the Hamiltonian $H_{\Phi}$, which is free of odd
operators to order $(1/c^2)$ . We do not quote this well known
procedure \cite{pp8, pp16}. The final result is
\begin{equation}
H_{\Phi} \approx U+\beta(mc^2 + \frac{1}{2mc^2}O^2 -
\frac{1}{8m^3c^6}O^4) -\frac{1}{8m^2c^4}[O,[O,U]]\;\;,
\end{equation}
where the brackets symbolize commutators.

Evaluating the above quantities we will use a 4x4 spin vector
operator \cite{pp11, pp16}
\begin{equation}
\bm{\Sigma}=\left| \begin{array}{cc} \bm{\sigma} & 0 \\ 0 &
\bm{\sigma}
\end{array} \right|\;\;.
\end{equation}
It has the following property for any two vectors \textbf{C} and
\textbf{D}
\begin{equation}
({\bm \alpha} \cdot {\bm C})({\bm \alpha} \cdot {\bm D})={\bm
C}\cdot {\bm D} + i\bm{\Sigma}\cdot (\bm{C} \times \bm{D})
\end{equation}
 Using this identity we obtain
\begin{equation}
\frac{1}{2mc^2}O^2 = \frac{1}{2m}\Pi^2 + \mu \bm{\Sigma} \cdot
\bm{B} \;\;,
\end{equation}
where $\mu=e\hbar/2m$ is the Bohr magneton and $\bf{B}={\bm
\nabla} \times \bf{A}$ is a magnetic field. Clearly, the operator
$O^4$ in Eq. (3) is just the square of $O^2$ given above.

The last term in Eq. (3) is calculated in two steps
\begin{equation}
[O,U]=[c\; \bm{\alpha} \cdot \bm {\Pi}, U]=c\hbar ei
(\bm{\alpha}\cdot{{\bm E}}) \;\;,
\end{equation}
where ${{\bm E}} = -\bm{\nabla}V$ is an electric field. To
evaluate the final commutator in Eq. (3) one needs two properties
of the spin vector operator $\bm{\Sigma}$, which follow directly
from the corresponding properties of $\bm{\sigma}$. We have
\begin{equation}
\alpha_k \alpha_j=\delta_{kj}+i\varepsilon_{kjl}\Sigma_l \;\;,
\end{equation}

\begin{equation}
[\alpha_k, \alpha_j]=-2i\varepsilon_{jkl}\Sigma_l \;\;,
\end{equation}
where $\delta_{kj}$ is the Kronecker delta and $\varepsilon_{jkl}$
is the antisymmetric unit tensor. The summation convention over
repeated subscripts is employed. Using Eqs (8) and (9) we
calculate
$$
[\bm{\alpha}\cdot \bm{\Pi}, \bm{\alpha}\cdot {\bm E}]=\alpha_k
\alpha_j[\Pi_k, {E}_j]+[\alpha_k, \alpha_j]{E}_j\Pi_k
$$
$$
=(\delta_{kj} + i\varepsilon_{kjl}\Sigma_l)\frac{\hbar \partial
{E}_j}{i
\partial x_k} -2i\varepsilon_{jkl}\Sigma_l{E}_j\Pi_k
$$
\begin{equation}
=-i\hbar(\bm{\nabla} \cdot {\bm E})+\hbar\bm{\Sigma}\cdot(\bm
{\nabla} \times {\bm E})-2i\bm{\Sigma}\cdot({\bm E} \times
\bm{\Pi})\;\;.
\end{equation}
If the scalar potential $V(\bf{r})$ has continuous first
derivatives, then $\bm{\nabla}\times {\bm E} =
-\bm{\nabla}\times\bm{\nabla}V = 0$, and the second term vanishes.

There exists an alternative way to evaluate the above commutator
using directly Eq. (5)
\begin{equation}
[\bm{\alpha}\cdot \bm{\Pi}, \bm{\alpha}\cdot {\bm
E}]=-i\hbar(\bm{\nabla} \cdot {\bm E})+i\bm{\Sigma}\cdot(\bm {\Pi}
\times {\bm E}-{\bm E} \times \bm{\Pi})\;\;.
\end{equation}
Both above forms are clearly equivalent. In agreement with the
common practice we use the gradient sign $\bm{\nabla}$ to
emphasize that the differentiation concerns the electric field
alone and not the wave function.

The Hamiltonian (3) factorizes into two 2x2 blocks for upper and
lower components of the wave function. For the upper block (
positive energies ) $\beta$  is to be replaced by +1 and
$\bm{\Sigma}$ by $\bm{\sigma}$ [cf. Eq. (4)]. This finally gives
for an electron with positive energies the following Hamiltonian
$$
H_{\Phi}=mc^2-eV+\frac{1}{2m}\Pi^2+\mu\bm{\sigma}\cdot\bm{B} +
$$
\begin{equation}
-\frac{1}{2mc^2}(\frac{1}{2m}\Pi^2 +
\mu\;\bm{\sigma}\cdot\bm{B})^2
+\frac{e\hbar}{4m^2c^2}\bm{\sigma}\cdot({\bm E}\times\bm{\Pi})
+\frac{e\hbar^2}{8m^2c^2}(\bm{\nabla}\cdot {\bm E})
\end{equation}

This is the main result of our paper. Now we briefly discuss its
physical content.

The fourth Pauli term is nonrelativistic although it results from
the relativistic Dirac equation. As shown by Huang \cite{pp21} and
Feshbach and Villars \cite{pp22}, this term is related to
electron's Zitterbewegung.

The sixth term represents the spin-orbit interaction. It is
written in the form calculated in Eq. (10) accounting for
$\bm{\nabla}\times {\bm E}=0$. The alternative form is [ see Eq.
(11)]
\begin{equation}
H_{so}=\frac{e\hbar}{8m^2c^2}\bm{\sigma} \cdot ({\bm E}\times\bm
{\Pi}-\bm{\Pi} \times {\bm E})\;\;.
\end{equation}
 In both forms the canonical momentum $\bm{\Pi}$ appears,
assuring the gauge invariance of the Hamiltonian (12).
Unfortunately, many well known references give noninvariant form
with $\bm{p}$. As demonstrated by Feshbach and Villars
\cite{pp22}, the spin-orbit term results from the linear
contribution $\Delta r$ to the electron displacement caused by the
Zitterbewegung.

The seventh term proportional to $\bm{\nabla} \cdot {\bm E}$ is
the Darwin term. It is commonly interpreted as coming from the
quadratic contribution $(\nabla r)^2$ to the electron displacement
caused by the Zitterbewegung.

Finally, we want to discuss the fifth term in the Hamiltonian
(12). Of all the enumerated references [1-20], only Corinaldesi
and Strocchi \cite{pp7}, Messiah \cite{pp17} and Hecht \cite{pp20}
quote the term $(\bm{\sigma}\cdot\bm{\Pi})^4$, but, since it is
nowhere separated into the orbital and spin parts and interpreted,
it merits attention. To facilitate the discussion, it is helpful
to consider first a simple situation of an electron in a constant
magnetic field \textbf{B}. In this case the Dirac equation has
exact solutions and the positive eigenenergies are \cite{pp23,
pp24}
\begin{equation}
{\epsilon}=[(mc^2)^2+2mc^2D(n,p_z,\pm)]^{1/2} \;\;,
\end{equation}
where
\begin{equation}
D(n,p_z,\pm)=\hbar\omega_c (n+\frac{1}{2}) + \frac{p_z^2}{2m}
\pm\mu B \;\;,
\end{equation}
in which $n =0,1,2,\ldots$ is the Landau quantum number and
$\omega_c=eB/m$ is the cyclotron frequency. Expanding the square
root for $2D \ll mc^2$, one obtains
\begin{equation}
{\epsilon}\approx mc^2+D-\frac{1}{2mc^2}D^2 \;\;.
\end{equation}
We can identify the above expression with the first, third, fourth
and fifth terms of the Hamiltonian (12) because, for the free
electron, the eigenvalue of the orbital term $\Pi^2/2m$ is
$\hbar\omega_c (n+1/2) + p^2_z/2m$. Thus the fifth term in Eq.
(12) corresponds to $(1/c)^2$ order in expansion of relativistic
energy (including the Pauli term $\mu\;\bm{\sigma}\cdot\bm{B}$).

If we keep $V=0$ but have otherwise arbitrary time independent
magnetic field ${\bf B}({\bf r})$, one can transform the Dirac
Hamiltonian for positive energies into the form ( see Case
\cite{pp25}, Eriksen and Kolsrud \cite{pp26})
\begin{equation}
H'=\beta mc^2[1+\frac{2}{mc^2}(\frac{\Pi^2}{2m} + \mu\;\bm{\sigma}
\cdot \bm{B})]^{1/2}\;\;.
\end{equation}

Expanding the square root and retaining the first three terms we
obtain exactly the corresponding terms in Eq. (12).

Equation (15) shows that the spin contribution to the fifth term
is not negligible in comparison to the orbital contribution. In
fact, for the $n=0$ Landau level the two contributions are exactly
equal. In other words, the spin splitting of electron energies is
equal to the orbital splitting.

Turning to the physical meaning of the term in question, it is
often stated that the $-(1/8m^3c^2)\Pi^4$ term in the $v^2/c^2$
expansion reflects relativistic increase of the electron mass with
increasing energy. As to the spin term, it was demonstrated by
Zawadzki \cite{pp27} that in the limit of vanishing magnetic
fields the spin magnetic moment of a free relativistic electron is
\begin{equation}
\mu({\epsilon})=\frac{e\hbar}{2m({\epsilon})}\;\;,
\end{equation}
where $m({\epsilon})={\epsilon}/c^2$ is the energy dependent
relativistic mass. Thus, for the electron at rest, ${\epsilon} =
mc^2$, the spin magnetic moment reduces to the Bohr magneton, but,
as the energy and the mass increase, the moment $\mu({\epsilon})$
decreases tending to zero. In the presence of magnetic field the
decrease of the magnetic moment means that the spin splitting of
the energy decreases with increasing energy. This decrease of the
spin splitting can be seen from Eqs (14) and (15): for a given
value of $B$ both the orbital and the spin splittings diminish as
$n$ grows. It is then clear that the appearance of the complete
fifth term in Eq. (12) which, as shown above, corresponds to the
expansion (16) or, more precisely, to the expansion of Eq. (17),
expresses not only relativistic increase of the electron mass but
also relativistic decrease of the spin magnetic moment.

Since the orbital and spin contributions to the fifth term
commute, one can perform the indicated squaring directly. In this
form the $v^2/c^2$ expansion of the Dirac equation, as given by
Eq. (12), contains two additional terms compared to the
expressions given in the literature.

In summary, we have critically examined the semirelativistic
expansion of the Dirac equation with scalar and vector potentials.
The complete, gauge invariant form of the expansion to order
$(1/c)^2$ is established. This form contains two additional terms,
as compared to different expressions given in original papers and
textbooks. It is demonstrated that the additional terms describe
relativistic decrease of the electron spin magnetic moment with
increasing electron energy.

\begin{acknowledgments}
I am pleased to thank Professor I. Bialynicki-Birula, Dr T.M.
Rusin and Dr P. Pfeffer for informative discussions. This work was
supported in part by The Polish Ministry of Sciences, Grant No
PBZ-MIN-008/PO3/2003.
\end{acknowledgments}

\end{document}